\documentclass[12pt]{article}
\usepackage{graphicx}
\usepackage{epsfig}

\begin{document}
\title{Incomplete information in scale-free networks}
\author{K. Hamacher\\
 Center for Theoretical Biological Physics\\
  University of California, San Diego\\
             La Jolla, CA 92093-0374, USA}

\maketitle


  \begin{abstract}
    We investigate the effect of incomplete information  on the  growth process of
    scale-free networks - a situation that occurs frequently
    e.g.~in real existing citation networks. Two models are proposed and
    solved analytically for the scaling behavior of the connectivity distribution.
    These models show a varying scaling exponent with respect to the model parameters but
    no break-down of scaling thus introducing the first models of
    scale-free networks in an environment of incomplete information.
    We compare to results from computer simulations which show a very good 
    agreement.
  \end{abstract}



  \newcommand{\kst}{\bar{k}(s,t)}
  \newcommand{\kut}{\bar{k}(u,t)}
  \newcommand{\pst}{\phi(s,t)}
  \newcommand{\Put}{\phi(u,t)}
  \newcommand{\pstk}{\phi_{<}(s,t)}
  \newcommand{\putk}{\phi_{<}(u,t)}
  \newcommand{\px}{\phi(x)}
  \newcommand{\upd}{d}
  \newcommand{\ab}{}
  \newcommand{\pxk}{\phi_{<}(x)}
  \newcommand{\kx}{\kappa(x)}

  \section{Introduction}
  Since the work on complex networks by Strogatz, Watts,
  Barab{\`a}si and Albert (see 
  \cite{Strogatz2001,Albert1999,Watts1998,Barabasi1999,Newman2003,sb,Albert2002,Dorogovtsev2002}) 
  many researchers from such distinct fields as statistical mechanics 
  \cite{dorogovtsev:027104,costa:066127,ramasco:036106,klemm2004,nr_b1},
  molecular biology \cite{jeong2000,jeong2001,nr_b3,Alm2003,Wagner2003,Amoutzias2004_2},
  ecology \cite{Mendoza1998}, physical chemistry \cite{Doye2002_2,Doye2002,Cancho2003}, 
  genetics e.g.~\cite{babu2004,Amoutzias2004,nr_b2,nr_b4,Conant2003} or social science
  \cite{bagrow2004,Ravid2004,newman:066133} have studied
  the emerging complex structure and the  behavior of networks in  their respective 
  field of research. 

  A special subset of scale-free non-equilibrium networks can emerge from a construction procedure in 
  which at each
  time step $t$ one vertex is added and connected to $m$ existing vertices   with preferential 
   linking. This preference
  is proportional to the number of already existing connections of that particular vertex. By definition the average number
  of connections remains constant $\bar{k}=2m$. 
  The distribution of the degree of connections $P(k)$ is of particular interest as it
  provides for the possibility to distinguish different classes. 
  One observes in scale-free-networks
  a behavior $P(k) \sim k^{-\gamma}$ that was first discussed by Simon  \cite{simon1953} and in the
  context of citations networks by Price \cite{price1976}. Extensions to 
  a more complex linking procedure \cite{PPredner2004} or more general linkage 
  properties \cite{PPredner2004_2}
  were recently discussed in detail.

  To study the evolution of the distribution
  the continuum approximation is often used. At time $t$  the average number of connections
   $\kst$ of a vertex created at time $s$  is
  in an undirected network \cite{Dorogovtsev2002,ph_do2}
  \begin{equation}
    \label{eq::approx}
    \frac{\partial\kst}{\partial t} = m\cdot \frac{\kst}{\int\limits_{0}^{t} \upd u\;\kut   }
  \end{equation}
    
   Bianconi and Barab{\`a}si  \cite{Bianconi2001} pointed to the effects of distributions
  of fitness of individuals to attract new connections. This can be already
   regarded as one prototypical example
  of incomplete information by interpreting the fitness in their model as an incomplete knowledge of all
  newer vertices about the {\em individual} properties or existence of the {\em present} vertices. 
  
  Mossa et al.\cite{mossa2002} showed that the power-law behavior
  might be truncated due to information filtering. In their case a newly attached vertex is only aware
  of a certain subset of the existing vertices. This subset is however chosen  randomly for 
  each vertex individually.
  Therefore the incomplete information has no global properties but is instead a local property.

  Here we want however to follow another route with two distinct 
  models to deal with the more interesting case that the incomplete information is
  attached to the new vertices individually and still global with respect to the whole network. 
  One model will mimic a generic and global effect that is present 
  in all real citation networks  while the other describes the
  influence of individual information unawareness. 

  \section{Growing nets and latency}

  For a newly created vertex the only relevant information is a list of existing vertices to connect to
  and their respective degrees.
  Incomplete information results in the ignorance of some of those vertices.
  This effect is here mediated via an 'awareness' function $\pst$ that makes the newly 
  connected vertex $t$
  aware of the  existing vertex $s$. 
  \[
  \label{eq::phi2}
  \pst := \left\{ \begin{array}{l@{\quad}l}  
    1 & s\mbox{ is known to } t \\ 0 & \mbox{else}
  \end{array} \right.
  \]

  Eq.~(\ref{eq::approx}) becomes then
  \begin{equation}
    \label{eq::phi}
    \frac{\partial\kst}{\partial t} = m \cdot \frac{\kst\pst}{\int\limits_{0}^{t} \upd u\;\kut\Put   }
  \end{equation}

  While there are many choices for $\pst$ we will propose one particular structure to resemble actual
  effects in citation networks. We will further set $m=1$ as this network property  
  has no influence on the scaling exponent $\gamma$ in the models.

  A newly created link might not be aware of the most recent created entities in the network.
  One encounters this situation actually very often: the author of a new WWW-page cannot be aware
  of other recently created pages that he would like to link to. Search engines do not
   provide for instantaneous
  listing of just created pages and therefore authors can find new pages  just by chance. 
  
  This incomplete information
  about the vertices of the network  results  in a selection of  'older' vertices for linking.
  We model this by the setting $\pstk := \Theta\left(c\cdot t - s \right)$ with 
  some constant $0<c\le 1$ and
  $\Theta(x)$ the Heaviside-step-function. Therefore a newly created vertex is only aware 
  of the oldest $c$ fraction of the existing vertices.

  While this seems to be at a first glance to strong of an assumption, this is realized in
  exponentially growing systems like the WWW: suppose from the current real-time $\tau$ we
  can not know new pages younger than some period $\tau_c$. 
  Then all existing pages that are capable of
  attracting a link  are taken from the  interval
  $\left[0;\tau-\tau_c\right]$. As the
   number of pages obeys however $\sim\exp\left(\alpha\cdot \tau\right)$
  this translates in the  $N$-notion to
  $\left[0;c\cdot N\left(\tau\right)\right]$. Here we have set $c:=\exp\left(-\alpha \tau_c\right)$.
  Recall that the time $t$ of eq.~(\ref{eq::approx}) is actually the number of vertices so
  that we actually draw a vertex from   $\left[0;c\cdot t\right]$.

  Rewriting the definition of $\pstk$ we get
  $\pstk = \Theta\left( c - s/t\right)$ and see that this function scales with respect to $s/t$. 
  We can then set $x=s/t$ and rewrite equation (\ref{eq::phi}) to (with $\kst = \kx$)
  \begin{equation}
    - x \frac{d\kx}{d x} =   \frac{\kx\pxk}{\int\limits_{0}^{1} \upd x \;\kx\pxk   }
    =   \frac{\kx\pxk}{\int\limits_{0}^{c} \upd x \;\kx   }
    =   \beta \cdot {\kx\pxk} \quad\mbox{with}\quad \kappa(1) = 1
  \end{equation}
  $\beta^{-1}$ is the integral in the denominator \cite{ph_do2}. The boundary condition $\kappa(1) = 1$
  reflects the fact, that upon creation a new node $s=t$ has only one connection.  The 
   solution  of this differential  equation is 
  \[
  \kx =  \left\{ \begin{array}{l@{\quad}l}  
    \tilde{c}\cdot x^{-\beta} &  0<t\le c \\  \mbox{const}=1 & t > c
  \end{array} \right.
  \]
  As the function has to be continuous we must have 
   $\kappa(1) {=} \kappa(c)$. We can derive the value for $\tilde{c}$ and
  arrive at
  \[
  \kx =  \left\{ \begin{array}{l@{\quad}l}  
    \left(\frac{x}{c}\right)^{-\beta} &  0<t\le c \\  1 & t > c
  \end{array} \right.
  \]

  The equation for $\beta$ is then
  \[
  \frac{1}{\beta} = \int\limits_{0}^{c} \kx \;\upd x = \frac{c}{1-\beta}   \quad
  \Longleftrightarrow \quad\beta = \frac{1}{1+c}
  \]

\begin{figure}
   \hspace*{-1cm}\begin{tabular}{cc}
    a) \includegraphics[scale=0.35]{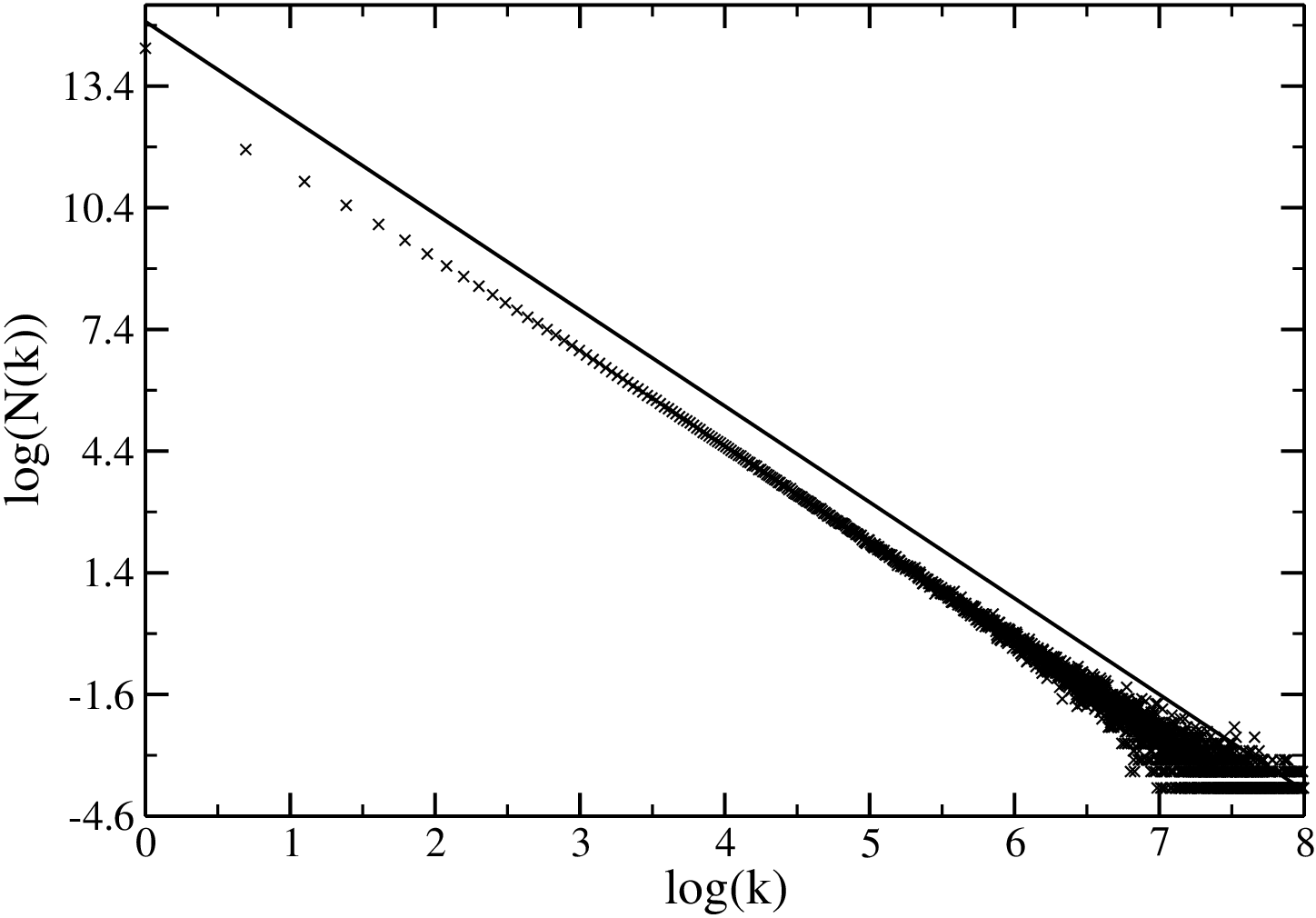} 
    b) \includegraphics[scale=0.35]{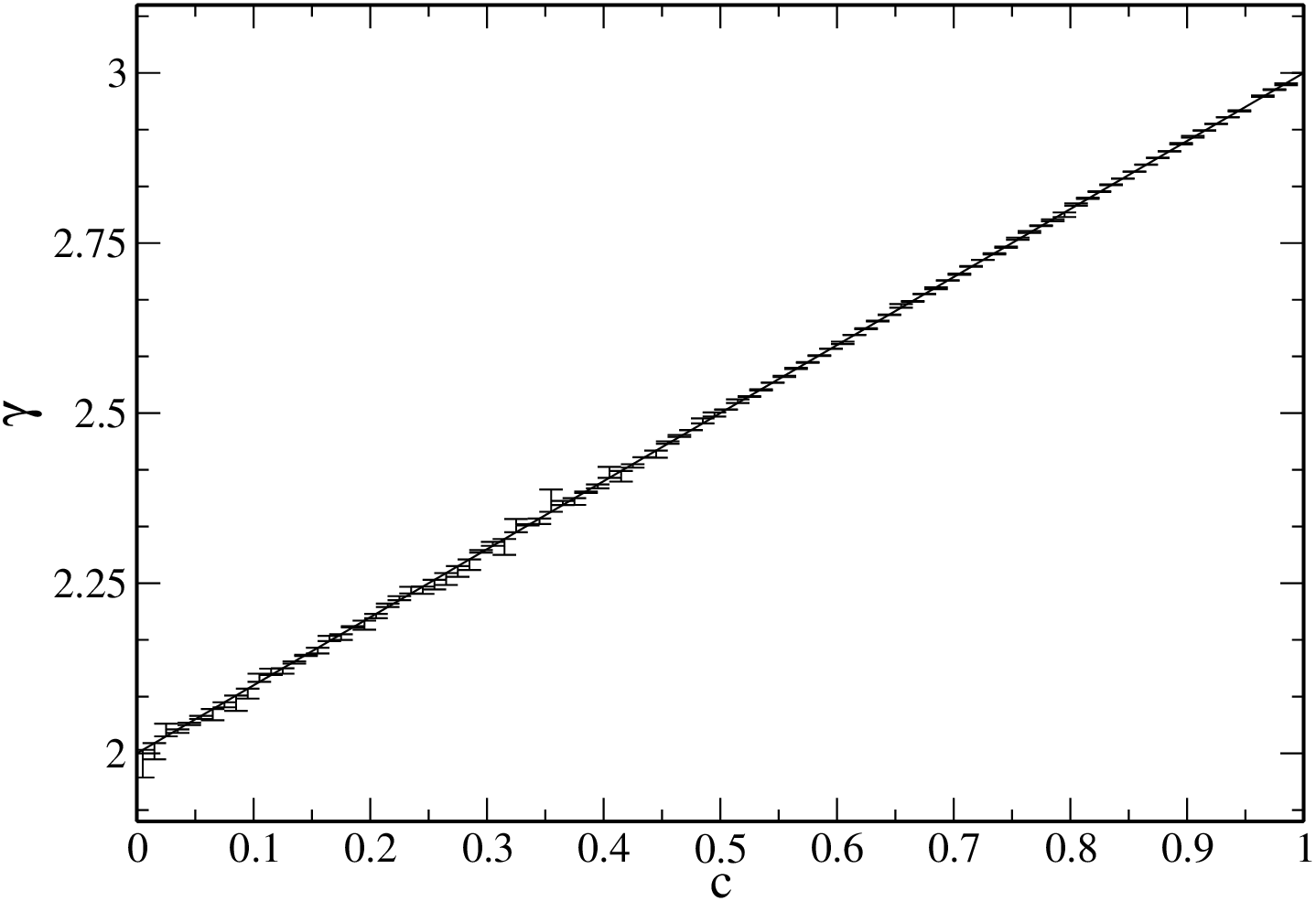} 
   \end{tabular}
    \caption{a) For $c=0.4$ with N=$10^6$ and sampled over $100$ replica we obtain
      a good fit to $N(k)=N\cdot p(k) \sim k^{-\gamma}$ with $\gamma_{fit}\approx 2.37$
      in the interval $\log(k)\in\left[2.2;5.5\right]$. The fitted curve was shifted for
       clarity.
      b) The exponent in the scaling law $P(k) \sim k^{-\gamma}$ as a function of the 
      information awareness $c$ in simulated 
      networks.  The straight line is the function
      $\gamma=2+c$. For each point  $600$ independent networks of  $10,000$ vertices each
      were sampled and the best fit was used. The error bars stem from the
      fitting procedure of the cumulative distribution \cite{ph_do2}. 
   }
    \label{fig::C1}
  \end{figure}

  By scaling arguments one can prove that 
  in general the relation $\gamma=1+1/\beta$ holds \cite{ph_do2,Dorogovtsev2000_2,Dorogovtsev2000}.
  Using this we conclude that here $\gamma = 2+c$. This
  is depicted in figure \ref{fig::C1} which compares this result
  with computer simulations.

\begin{figure}
    \includegraphics[scale=0.5]{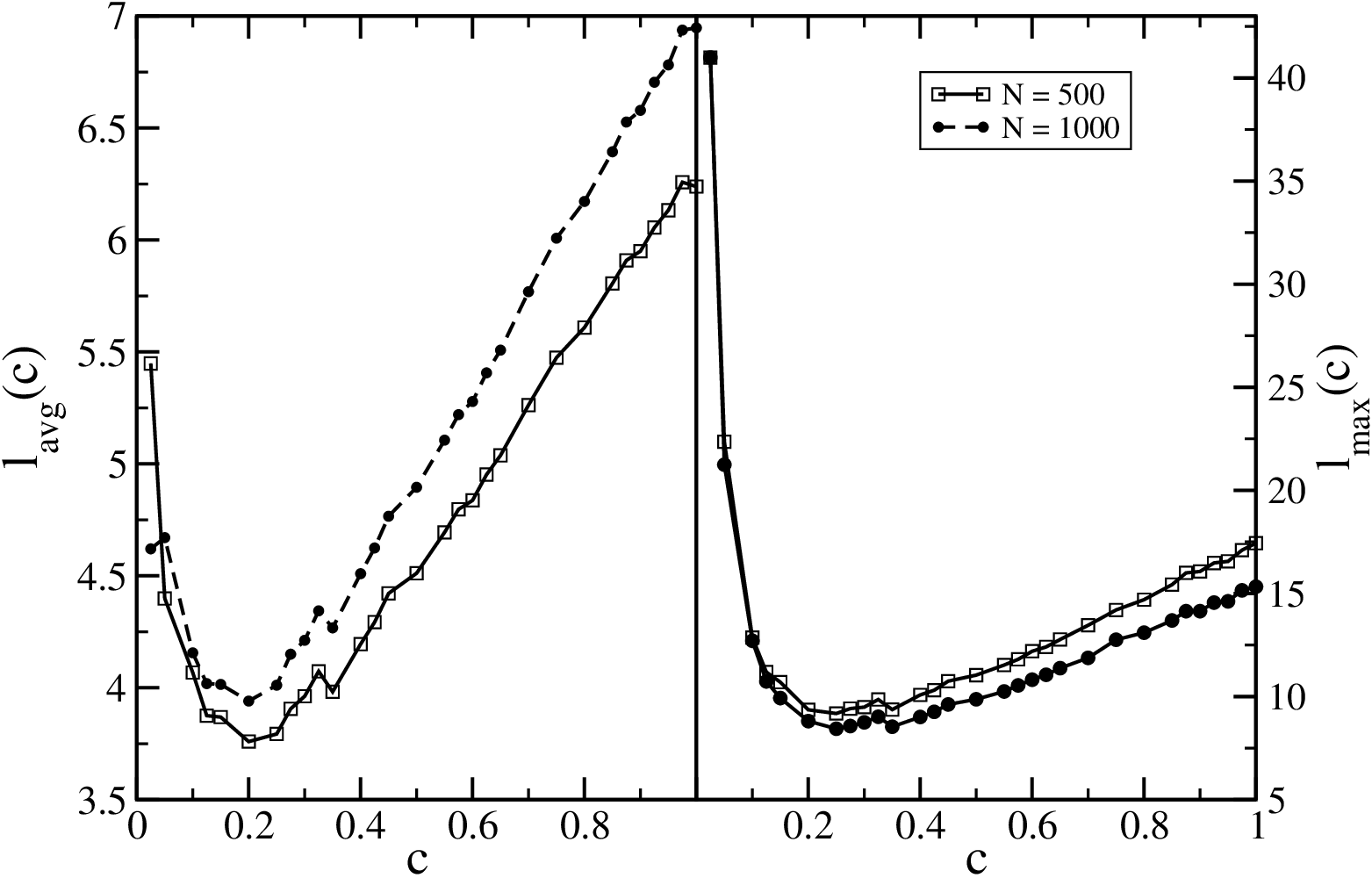}
    \caption{The average and the maximum of the minimal path lengths in the network grown with
     an information awareness of $c$. For every point we sampled over $100$ independent
     networks. The first $1/c$ vertices where initialized to form a chain, so giving
    no preferences to any of them.}
    \label{fig::SP}
  \end{figure}

  We can further analyze the emerging networks by investigating the shortest-paths in these
  networks. Figure \ref{fig::SP} shows the increase of both the average  $l_{avg}$ 
   and the maximum  $l_{max}$
  of the shortest path
   lengths  \cite{nr_pajek}  
  in independently created networks with increasing $c$ above $c\approx 0.2$.

  A smaller $c\in \left[0.2;1.0\right]$ reduces the number of available vertices in the growth process
  and the network gets more dense: the shortest path lengths get smaller and the probability
  of vertices with larger number of connections bigger - as can be seen from 
  the previous result $\gamma=2+c$. 

  For very small $c<0.2$ we see however an increase in
  $l_{avg}$ and $l_{max}$. This is an artifact due to our initial setting of the first
  $1/c$ vertices to form a chain. While the chain guarantees that none of the first
  vertices is preferred to another initial vertex, the path lengths are 
  largely influenced now. For the smallest $c=0.025$ used here we get an $l_{max}=40.99\approx 1/c$.
  Here the chain length dominates the path lengths.
 
  Suppose that WWW-pages are the vertices in this scenario and links are the edges of the network.
  This model then takes into account the time that e.g.~search engines need to encounter new 
  web-pages and
  make the general public aware of those sites. 
  The anonymity of a vertex $s$ is healed over time by sliding into the focus of new vertices as 
  soon as $s<c\cdot t$. 
  Here the incomplete information refers to the knowledge of {\em every} new vertex uniformly.

  There is also the opposite scenario in which some vertices are aware of the full information (that
  is the number of connections all the existing vertices possess) and others are just 
  ignorant and connect with
  equal probability to any of the existing ones. Here the incompleteness of information is restricted to 
  a subset of
  individuals:


  \section{Growing nets and partial ignorance}
\begin{figure}
    \includegraphics[scale=0.5]{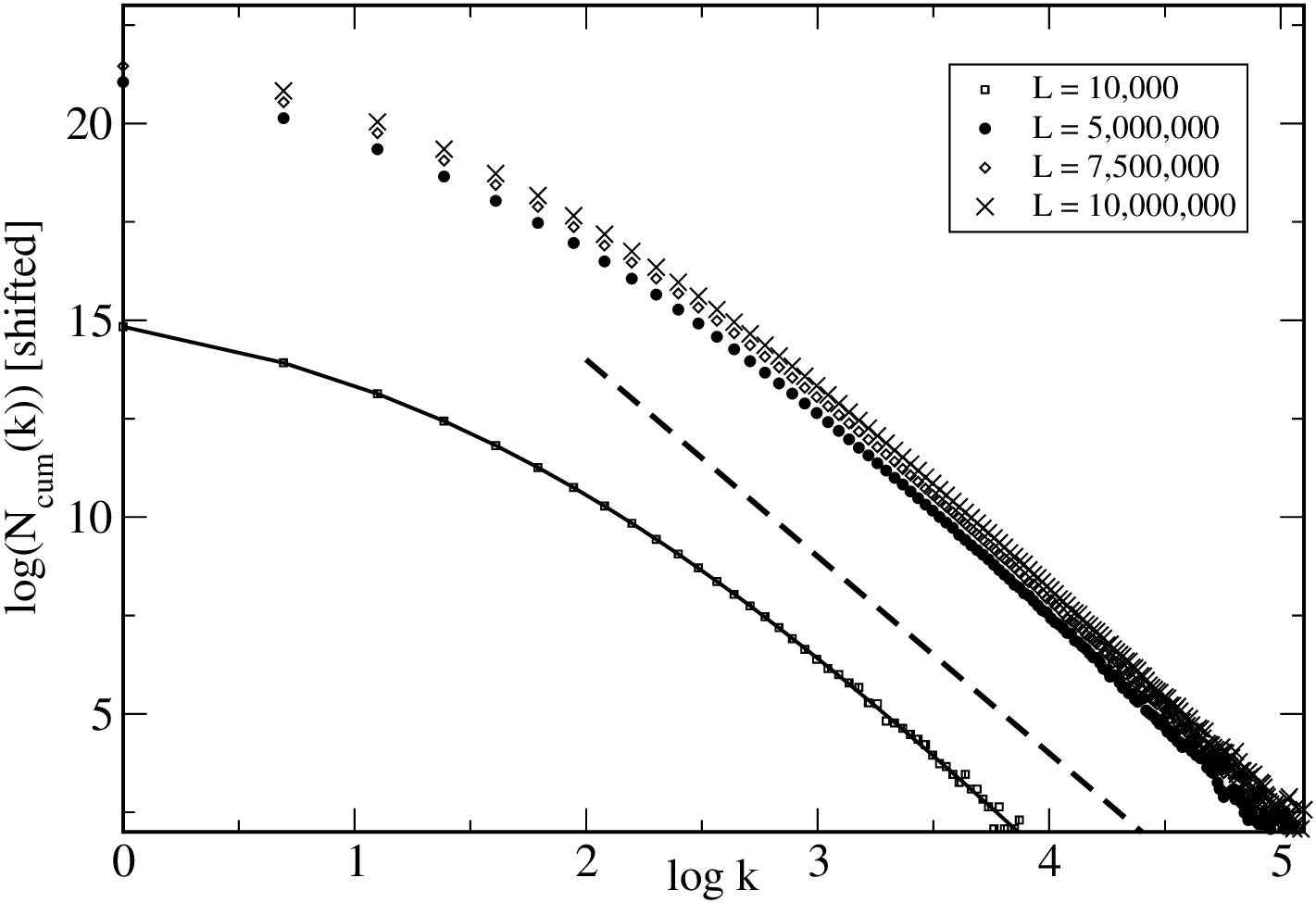}
    \caption{The cumulative number of connections $N_{\ab{cum}}(k)=\sum\limits_{k'=k}^{\infty} N(k')$ 
      of degree $k$ in networks of size $L$  averaged over
      $500$ independent runs. The data was shifted for a better overview.
      The straight line through the data of $L=10,000$ is the derived
      result of eq.~(\ref{eq::p2}) with $p=0.4$ while the broken line indicates 
      an asymptotic power-law for the   cumulative number of a distribution 
            with $\gamma = 1+2/p = 6$.}
    \label{fig::Np}
  \end{figure}

  Suppose that a newly added vertex is with some probability $p$ aware of all the connectivities of 
   the other vertices. In this
  case it is attached with preferential linking described above. With probability $1-p$ it 
  is connected without preference. 
  We want to deduce the effect on the connectivity distribution from the master equation for the
  average number of connections of degree $k$ at time $t$
  \begin{eqnarray}
    \label{eq::p}
    N(k,t+1) &=& p \cdot \left[ N(k,t) + \frac{k-1}{t\bar{k}} N(k-1,t) -
     \frac{k}{t\bar{k}}N(k,t)\right] \nonumber\\
    && + (1-p)\cdot \left[ N(k,t) + \frac{1}{t} N(k-1,t) - \frac{1}{t}N(k,t)\right] + \delta_{k,1}
  \end{eqnarray}
  Here $\bar{k}=2$ is the average degree of each vertex. The first term describes the 
  preferential linking with its 
  in- and outflow while the second term provides for the additional connections or loss thereof 
  with equal probability.
  Notice that there are currently $t$ vertices in the network, so $1/t$ is the probability of hitting 
  any one of those.
  The third term is finally responsible for the newly added 'guy'. By changing to  continuous 
   time we get from eq.~(\ref{eq::p})
  \[
  \frac{\partial}{\partial t} (t\cdot P(k,t)) = \left[ p\frac{k-1}{\bar{k}}+1-p\right]P(k-1,t)
  - \left[ p\frac{k}{\bar{k}}+1-p\right]P(k-1,t) + \delta_{k,1}
  \]
  where $P(k,t) = N(k,t)/t$ is the density of vertices with degree $k$
  at time $t$. 
  We can now solve for the stationary distribution for $t\rightarrow\infty$.
  We arrive at the recursion
  \[
  P(k) = \frac{pk-p+\bar{k}-p\bar{k}}{pk+2\bar{k}-p\bar{k}} P(k-1) +
           \frac{\bar{k}}{pk+2\bar{k}-p\bar{k}} \delta_{k,1}
  \]
  This is further written as
  \begin{equation}
    \label{eq::p2}
    P(k) =    \frac{2}{4-p} \cdot \frac{\Gamma\left(\frac{4}{p}\right)\Gamma\left(\frac{2}{p}-2+k\right)
                                      }{
                                \Gamma\left(\frac{2}{p}-1\right)\Gamma\left(\frac{4}{p}-1+k\right)
  } 
  \end{equation}
  Using the relationship 
  \[
  \frac{\Gamma \left(k+a\right)}{\Gamma\left(k+b\right)} = k^{a-b} \left[ 1  + O\left(k^{-1}\right) \right]
  \]
  for large $k$ \cite{abramowitz6147} we 
  conclude that $\gamma = 1 + \frac{2}{p}$ for large $k$ with a divergence for the 
   exponent when approaching $p=0$ as in this case we
  have no preferential linking at all. In this case the starting master equation (\ref{eq::p}) 
  leads correctly to the
  Poisson-distribution $P_{p=0}(k) = 2^{-k}$. The diverging behavior of the exponent was for instance 
  also found by Krapivsky and Redner
  \cite{krapivsky2001} in their treatment of growing networks with redirection.
  Figure \ref{fig::Np} shows the results from computer experiments for this model. 
  The smaller the $p$ the
  more difficult it is to see any indication of the power-law.

%

  \section{Conclusion}  
  In this paper we developed two distinct models to describe the effect of 1) global incomplete
  information caused by penetration rates while constructing a citation network and 2)
  local incomplete information of individual vertices that are attached with a probability of
  'non-knowledge'. We derived the scaling behavior of the degree distribution for large
  degrees in both cases and compared this to computer experiments. 
  Both models approach the analytic value  \cite{Barabasi1999} of $\gamma=3$ when reaching full information.
  The incomplete information in the two models does not destroy the scale-free-behavior of the systems
  while Mossa et al.\cite{mossa2002} found  a cross-over from scale-free-behavior to an exponential
  in another  model which takes information into account.  By comparison one can see the 
	influence incomplete  information may have on the global structure of growing networks.

  We will work out particulars on real-world-networks and the influence of incomplete information
  in a forthcoming study. 
  
  \section*{Acknowledgments}
  KH is supported through a Liebig-Fellowship of the Fonds der chemischen Industrie. 
  Computational resources were provided under a
  grant of the Howard Hughes Medical Institute and 
  by the NSF-sponsored Center for Theoretical Biological Physics 
  (grant numbers PHY-0216576 and 0225630).\\
  I thank S.~Redner and S.~Mossa for  bringing ref.~\cite{krapivsky2001} and \cite{mossa2002}  
   respectively to my attention.
  Stimulating discussions with and comments from C. Gros, J.A. McCammon, and T. Hwa 
  are gratefully acknowledged. 


\begin{thebibliography}{10}

\bibitem{Strogatz2001}
{ Strogatz S},
\newblock  2001 {\em Nature} {\bf 410} 268

\bibitem{Albert1999}
 Albert R,  Jeong H and  Barab{\`a}si A-L,
\newblock 1999 {\em Nature} {\bf 401} 130

\bibitem{Watts1998}
 Watts DJ and  Strogatz SH, 1998
\newblock {\em Nature} {\bf 393} 440

\bibitem{Barabasi1999}
 Barab{\`a}si A-L and  Albert R, 1999
\newblock {\em Science} {\bf 286} 509

\bibitem{Newman2003}
 Newman MEJ, 2003
\newblock {\em SIAM Review} {\bf 45} 167

\bibitem{sb}
{  Bornholdt S and  Schuster HG}, 2002
\newblock {\em {Handbook of Graphs and Networks}}
\newblock (Weinheim: Wiley-Vch)

\bibitem{Albert2002}
 Albert R and Barab{\`a}si  A-L, 2002
\newblock {\em Rev. Mod. Phys.} {\bf 74} 47

\bibitem{Dorogovtsev2002}
{  Dorogovtsev SN and  Mendes JFF }, 2002
\newblock {\em { Advances In Physics}} {\bf 51} 1079

\bibitem{dorogovtsev:027104}
 Dorogovtsev SN, 2004
\newblock {\em Phys. Rev. E} {\bf 69} 027104

\bibitem{costa:066127}
 da~Fontoura~Costa L, 2004
\newblock {\em Phys. Rev. E} {\bf 69} 066127

\bibitem{ramasco:036106}
 Ramasco JJ,  Dorogovtsev SN and Pastor-Satorras R, 2004
\newblock {\em Phys. Rev. E} {\bf 70} 036106

\bibitem{klemm2004}
Klemm K and Bornholdt S, 2004
\newblock {\em preprint q-bio/0409022}

\bibitem{nr_b1}
Reichardt J and Bornholdt S, 2004
\newblock {\em Phys. Rev. Lett.} {\bf 93} 218701

\bibitem{jeong2000}
Jeong H, Tombor B, Albert R,  Oltvai ZN and Barab{\`a}si AL, 2000
\newblock {\em Nature}, {\bf 407} 651

\bibitem{jeong2001}
Jeong H,  Mason SP,  Barab{\`a}si AL and  Oltvai ZN, 2001
\newblock {\em Nature} {\bf 411} 41

\bibitem{nr_b3}
Bornholdt S, 2001
\newblock {\em Biol. Chem.} {\bf 382} 1289

\bibitem{Alm2003}
Alm E and Arkin AP, 2003
\newblock {\em Curr. Op.  Struct. Biol.} {\bf 13} 193

\bibitem{Wagner2003}
 Wagner A, 2003
\newblock {\em Proc. R. Soc. Lond. B} {\bf 270} 457

\bibitem{Amoutzias2004_2}
 Amoutzias GD,  Robertson DL and Bornberg-Bauer E, 2004
\newblock {\em Comp. and Funct. Gen.} {\bf 5} 79

\bibitem{Mendoza1998}
 Mendoza L and Alvarez-Buylla ER, 1998
\newblock {\em J. theor. Biol.} {\bf 193} 307

\bibitem{Doye2002_2}
 Doye JPK and Wales DJ, 2002
\newblock {\em J. Chem. Phys.} {\bf 116} 3777

\bibitem{Doye2002}
 Doye JPK, 2002
\newblock {\em Phys. Rev. Lett.} {\bf 88} 238701

\bibitem{Cancho2003}
 Cancho RF and Sol{\`e} RV, 2003
\newblock {Optimization in Complex Networks}.
\newblock In { R. Pastor-Satorras, M. Rubi and A. Diaz-Guilera} (eds.), {\em
  {Statistical Mechanics of Complex Networks, Lecture Notes in Physics}}, 
  114--125, (Berlin: Springer)

\bibitem{babu2004}
 Babu MM, Luscombe NM, Aravind L,  Gerstein M and  Teichmann SA, 2004
\newblock {\em Curr. Op.  Struct. Biol.} {\bf 14} 283

\bibitem{Amoutzias2004}
 Amoputzias GD,  Robertson DL, Oliver SG and 
  Bornberg-Bauer E, 2004
\newblock {\em EMBO reports} {\bf 5} 1

\bibitem{nr_b2}
Bornholdt S and Sneppen K, 1998
\newblock {\em Phys. Rev. Lett.} {\bf 81} 236

\bibitem{nr_b4}
Klemm T and Bornholdt S, 2004
\newblock {\em Preprint q-bio/0409022}

\bibitem{Conant2003}
 Conant GC and Wagner A, 2003
\newblock {\em Nature Genetics} {\bf 34} 264

\bibitem{bagrow2004}
 Bagrow JP, Rozenfeld HD,  Bollt EM and ben Avraham D, 2004
\newblock {\em {Europhys. Lett.}} {\bf 67} 511

\bibitem{Ravid2004}
 Ravid G and Rafaeli S, 2004
\newblock {\em {First Monday}} {\bf 9}
\newblock {URL: http://firstmonday.org/issues/issue9\_9/ravid/index.html }

\bibitem{newman:066133}
 Newman MEJ,
\newblock {\em Phys. Rev. E} {\bf 69} 066133

\bibitem{simon1953}
 Simon RA, 1953
\newblock {\em Biometrika} {\bf 42} 425

\bibitem{price1976}
 de~Solla~Price D, 1976
\newblock {\em {J. A. Soc. Inf. Sci.}} {\bf  27} 292

\bibitem{PPredner2004}
 Krapivsky PL and Redner S, 2004
\newblock {\em Preprint cond-mat/0410379}

\bibitem{PPredner2004_2}
Almaas E, Krapivsky PL and Redner S, 2004
\newblock {\em Preprint cond-mat/0408295}

\bibitem{ph_do2}
 Dorogovtsev SN and  Mendes JFF, 2003
\newblock {\em {Evolution of Networks}}.
\newblock (Oxford: Oxford)

\bibitem{Bianconi2001}
Bianconi G and  Barab{\`a}si AL, 2001
\newblock {\em {Europhys. Lett.}} {\bf 54} 436

\bibitem{mossa2002}
 Mossa S,  Barth{\`e}l{\`e}my M,  Stanley HE and 
  Amaral  LA.
\newblock {\em Phys. Rev. Lett.} {\bf 88} 138701

\bibitem{Dorogovtsev2000_2}
 Mendes JFF, Dorogovtsev SN and  Samukhin AN, 2000
\newblock {\em Phys. Rev. Lett.} {\bf 85} 4633

\bibitem{Dorogovtsev2000}
 Dorogovtsev SN and Mendes JFF, 2000
\newblock {\em Phys. Rev. E} {\bf  62} 1842




\bibitem{nr_pajek}
Batagelj V and  Mrvar A, 1998
 {\em Connections} {\bf 21} 47

\bibitem{abramowitz6147}
 Abramowitz M and  Stegun IA, 1965
\newblock {\em Handbook of Mathematical Functions}.
\newblock (New York: Dover)

\bibitem{krapivsky2001}
{ Krapivsky PL and Redner S}, 2001
\newblock {\em Phys. Rev. E} {\bf 63} 066123

\end{thebibliography}


\end{document}